\documentclass[twocolumn,showpacs,preprintnumbers,amsmath,amssymb,floatfix]{revtex4}

\usepackage{graphicx}
\usepackage{dcolumn}
\usepackage{bm}
\begin{document}
\title{Nucleation and growth in one dimension,\\ part I: The generalized Kolmogorov-Johnson-Mehl-Avrami model}
\author{Suckjoon Jun}
\altaffiliation{Present address:  FOM Institute for Atomic and Molecular Physics (AMOLF), P.O.Box 41883, 1009 DB Amsterdam, The Netherlands}
\author{Haiyang Zhang}
\author{John Bechhoefer} 
\email[email: ]{johnb@sfu.ca}
\affiliation{Department of Physics, Simon Fraser University, Burnaby, B.C., V5A 1S6, Canada}
\date{\today}

\begin{abstract}
Motivated by a recent application of the Kolmogorov-Johnson-Mehl-Avrami (KJMA) model to the study of DNA replication, we consider the one-dimensional version of this model.  We generalize previous work to the case where the nucleation rate is an arbitrary function $I(t)$ and obtain analytical results for the time-dependent distributions of various quantities (such as the island distribution).  We also present improved computer simulation algorithms to study the 1D KJMA model.  The analytical results and simulations are in excellent agreement.
\end{abstract}


\maketitle

\section{Introduction}

Consider a tray of water that at time $t=0$ is put into a freezer.  A short while later, the water is all frozen.  One may thus ask, ``What fraction $f(t)$ of water is frozen at time $t \geq 0$?"  In the 1930s, several scientists independently derived a stochastic model that could predict the form of $f(t)$, which experimentally is a sigmoidal curve.  The ``Kolmogorov-Johnson-Mehl-Avrami" (KJMA) model~\cite{Kolmo, Johnson-Mehl, Avrami} has since been widely used by metallurgists and other materials scientists to analyze phase transition kinetics~\cite{Christian}.  In addition, the model has been applied to a wide range of other problems, from crystallization kinetics of lipids~\cite{Yang}, polymers~\cite{Huang}, the analysis of depositions in surface science~\cite{Fanfoni1}, to ecological systems~\cite{TomCaraco} and even to cosmology~\cite{Kampfer}.  For further examples, applications, and the history of the theory, see the reviews by Evans~\cite{JWEvans},  Fanfoni and Tomellini~\cite{Fanfoni1}, and Ramos {\it et al.}~\cite{Ramos}.

In the KJMA model, freezing kinetics result from three simultaneous processes: 1) nucleation of solid domains (``islands"); 2) growth of existing islands; and 3) coalescence, which occurs when two expanding islands merge.  In the simplest form of KJMA, islands nucleate anywhere in the liquid areas (``holes"), with equal probability for all spatial locations (``homogeneous nucleation").  Once an island has been nucleated, it grows out as a sphere at constant velocity $v$.  (The assumption of constant $v$ is usually a good one as long as temperature is held constant, but real shapes are far from spherical.  In water, for example, the islands are snowflakes; in general, the shape is a mixture of dendritic and faceted forms.  The effect of island shape -- not relevant to the one-dimensional version of KJMA studied here -- is discussed extensively in~\cite{Christian}.)  When two islands impinge, growth ceases at the point of contact, while continuing elsewhere.  KJMA used elementary methods, reviewed below, to calculate quantities such as $f(t)$.  Later researchers have revisited and refined KJMA's methods to take into account various effects, such as finite system size and inhomogeneities in growth and nucleation rates~\cite{Orihara, Cahn}.

Although most of the applications of the KJMA model have been to the study of phase transformations in three-dimensional systems, similar ideas have been applied to a wide range of one-dimensional problems, such as R\'{e}nyi's car-parking problem~\cite{Renyi} and the coarsening of long parallel droplets~\cite{Derrida}.  Recently, we have shown that the one-dimensional KJMA model can also be used to describe DNA replication in higher organisms~\cite{Herrick}.  Briefly, in higher organisms (eukaryotes), DNA replication is initiated at multiple origins throughout the genome.  A replicated domain then grows symmetrically with velocity $v$ away from the replication origin.  Domains that impinge coalesce.  And finally, each base in the genome is replicated only once per cell cycle.  Thus, if one views replicated regions as ``solid," unreplicated ones as ``liquid," and the initiation of replication origins as ``nucleation," all of the essential ingredients of the KJMA model are present.
\begin{figure}[!t]
\centering
\includegraphics[width=3.4in]{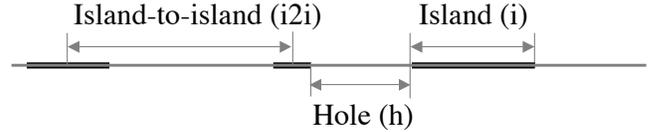}
\caption{Definitions.  In the KJMA model, a hole is the liquid domain between the growing solid domains (island).  The island-to-island is defined as the distance between the centers of two adjacent islands.}
\label{fig:definition}
\end{figure}
The purpose of the present two papers, then, is as follows:  Here, in paper I, we discuss how to generalize the KJMA model for biological application.  In particular, we consider the problem of arbitrarily varying origin initiation rate (equivalent to arbitrarily varying nucleation rate in freezing processes).  Then, in paper II, we discuss a number of subtle but generic issues that arise in the application of the KJMA model to DNA replication.  The most important of these is that the method of analysis runs backward from the usual one.  Normally, one starts from a known nucleation rate (determined by temperature, mostly) and tries to deduce properties of the crystallization kinetics.  In the biological experiments, the reverse is required: from measurements of statistics associated with replication, one wants to deduce the initiation rate $I(t)$.  This problem, along with others relating to inevitable experimental limitations, merits separate consideration.

In the mid-1980s, Sekimoto showed that the analysis of the KJMA model could be pushed much further if growth occurs in only one spatial dimension~\cite{Sekimoto}.  Sekimoto used methods from non-equilibrium statistical physics to describe the detailed statistics of domain sizes and spacings, as defined in Fig.~\ref{fig:definition}.  In particular, he studied the time evolution of domain statistics by solving Fokker-Planck-type equations for island and hole distributions, for constant nucleation rate $I(t)$=const. His approach has since been revisited by others (e.g.~\cite{Ben-Naim1996}).

Below, we extend Sekimoto's approach to the case of an arbitrary nucleation rate $I(t)$.  As mentioned above, this case is relevant to the kinetics of DNA replication in eukaryotes.  We also present two algorithms to simulate 1D nucleation and growth processes that are both much faster than more standard lattice methods~\cite{Krug}.

\section{Theory}

\subsection{\label{sec:f}Island fraction $f(t)$}
\begin{figure}[!t]
\centering
\includegraphics[width=3.4in]{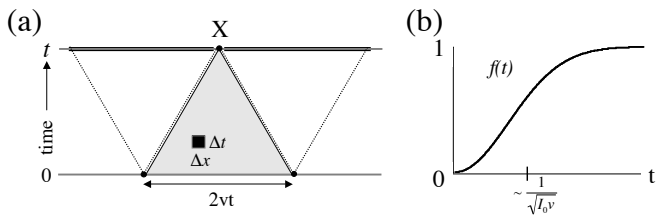}
\caption{Kolmogorov's method.  (a) Spacetime diagram.  In the small square box, the probability of nucleation is $I_0 \Delta x \Delta t$, where $I_0$ is the nucleation rate.  In order for the point $X$ to remain uncovered by islands, there should be no nucleation in the shaded triangle in spacetime. (b) Kinetic curve for constant nucleation rate $I_0$: $f(t) = 1- \exp(-I_0 v t^2)$. }
\label{fig:f(t)}
\end{figure}
We begin with the calculation of $f(t)$, the fraction of islands at time $t$ in a one-dimensional system. We write as $f(t) = 1- S(t)$, where $S(t)$ is the fraction of the system uncovered by islands (i.e., the hole fraction).  In other words, $S(t)$ is the probability for an arbitrary point $X$ at time $t$ to remain uncovered.  If we view the evolution via a two-dimensional spacetime diagram [Fig.~\ref{fig:f(t)}(a)], we can calculate $S$ by noting that
\begin{eqnarray}
\label{eq:S(t)}
S(t) &=& \lim \limits_{\Delta x, \Delta t \rightarrow 0} \prod_{x, t \in \bigtriangleup} (1-I_0 \Delta x \Delta t)\nonumber\\
&=& \exp\bigg(-\iint_{x, t \in \bigtriangleup} I_0 dxdt\bigg)\\
&=& \exp(-I_0 v t^2).\nonumber
\end{eqnarray}

\noindent Therefore,
\begin{equation}
\label{eq:f(t)}
f(t) = 1-e^{-I_0 v t^2},
\end{equation}

\noindent which has a sigmoidal shape, as mentioned above [see Fig.~\ref{fig:f(t)}(b)].

We note that Kolmogorov's method can be straightforwardly applied to any spatial dimension $D$ for arbitrary time- and space-dependent nucleation rates $I(\vec{x}, t)$.  Similar ``time-cone" methods can yield $f(t)$ in the presence of complications such as finite system sizes~\cite{Orihara, Cahn}.  Unfortunately, this simple method cannot be used to calculate the distributions defined in Fig.~\ref{fig:definition}, except that it can partly help solve the time-evolution equation for the hole-size distribution (see below).

\subsection{\label{sec:rhoh}Hole-size distribution $\rho_h(x,t)$}

We define $\rho_h(x, t)$ as the density of holes of size $x$ at time $t$.  For a spatially homogeneous nucleation function $I(t)$, the density $\rho_h$ will also be spatially homogeneous (The hole size $x$ should not be confused with the genome spatial coordinate $X$).  The time evolution $\rho_h(x, t)$ then obeys
\begin{eqnarray}
\label{eq:rhoh}
\frac{\partial \rho_h(x,t)}{\partial t} &=& 2v \frac{\partial \rho_h(x,t)}{\partial x}-I(t) x \rho_h(x, t)\nonumber\\
&& +2I(t) \int_{x}^{\infty} \rho_h(y, t) dy,
\end{eqnarray}

\noindent where $v$ is the growth velocity of islands and $I(t)$ is the spatially homogeneous nucleation rate at time $t$~\cite{Sekimoto}.  The first term on the right-hand side describes the effects on $\rho_h(x,t)$ of domain growth in the absence of coalescence and nucleation.  The second term accounts for the annihilation of a hole of size $x$ by nucleation, while the last term represents the splitting of a hole larger than $x$ by nucleation.
Eq.~\ref{eq:rhoh} was solved by Sekimoto for $I(t)$=const., while Ben-Naim {\it et al.} derived a formal solution for arbitrary $I(t)$~\cite{Ben-Naim1998}.  Below, we show that the solution of Ben-Naim {\it et al.} can also be obtained directly by applying Kolmogorov's argument.
\begin{figure}[!t]
\centering
\includegraphics[width=3.4in]{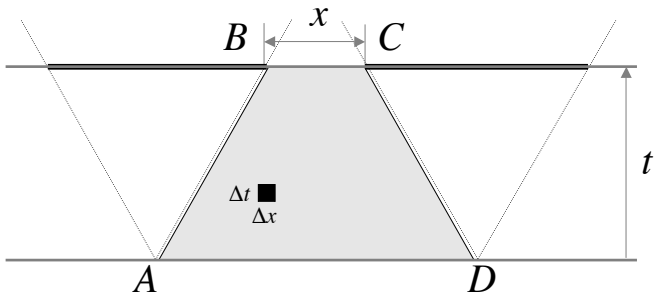}
\caption{Spacetime diagram.  The hole-size distribution $\rho_h(x,t)$ is proportional to the probability
$p_0(x,t)$ for no nucleation event occurs in the shaded parallelogram $ABCD$ (see text).}
\label{fig:kolmogorov}
\end{figure}
In Fig.~\ref{fig:kolmogorov}, we see a hole of size $x$ flanked by two islands.  In order for such holes to exist at time $t$, there should be no nucleation within the parallelogram $ABCD$ in the spacetime diagram.  Similar to the calculation of the hole fraction $S(t)$, we obtain the ``no nucleation" probability in the parallelogram as
\begin{eqnarray}
\label{eq:p0(t)}
p_0(t) &=& \lim \limits_{\Delta x, \Delta t \rightarrow 0} \prod_{x, t \in ABCD} [1-I(t) \Delta x \Delta t]\nonumber\\
&=& S(t) e^{-g(t) \cdot x},
\end{eqnarray}
\noindent where $g(t) = \int_0^t I(t') dt'$.  The domain density $n(t)$ and the hole fraction $S(t)$ are related by definition as follows:
\begin{eqnarray}
\label{eq:n}
n(t) &=& \int_{0}^{\infty} \rho_h(x, t) dx\\
S(t) &=& \int_{0}^{\infty} x \rho_h(x, t) dx.
\end{eqnarray}
\noindent Since the hole-size distribution $\rho_h(x,t)$ is proportional to $p_0(x,t)$, we can write $\rho_h(x,t) = c(t) \cdot p_0(x,t)$.  By integrating this equation and using Eq.~\ref{eq:n}, we obtain $c(t) = n(t) \cdot g(t) / S(t)$.  Putting this back into Eq.~\ref{eq:rhoh}, we obtain an equation for $n(t)$:
\begin{equation}
\label{eq:dndt}
\frac{1}{n(t)}\frac{\partial n(t)}{\partial t} = -2v \cdot g(t) + \frac{I(t)}{g(t)}.
\end{equation}

\noindent This is a first-order linear equation and can be solved exactly.  Using the boundary condition $n(0)=1$, we solve Eqs.~\ref{eq:dndt} and \ref{eq:rhoh} to find
\begin{eqnarray}
\label{eq:sol_rhoh}
n(t) &=& g(t) \cdot e^{-2v \int_0^t {g(t')dt'}};\\
\rho_h(x, t) &=& g(t)^2 \cdot e^{-g(t)x - 2v \int_0^t {g(t')dt'}}.
\end{eqnarray}
\noindent These are just exponential functions of $x$, with decay constants that monotonically decrease as a function of time.

\subsection{\label{sec:rhoi}Island distribution $\rho_i(x,t)$}

In analogy to Eq.~\ref{eq:rhoh} and following~\cite{Sekimoto}, the time evolution of the island distribution $\rho_i(x, t)$ is governed by drift, creation, and annihilation terms, as follows:
\begin{widetext}
\begin{equation}
\label{eq:rhoi}
\frac{\partial \rho_i(x,t)}{\partial t} = -2v \frac{\partial \rho_i(x,t)}{\partial x} + I(t) S(t) \delta(x)+2v \frac{\rho_h(0, t)}{n(t)^2} \Bigg[\int_{0}^{x} \rho_i(x-y,t)\rho_i(y,t) dy - 2n(t) \rho_i(x,t)\Bigg].
\end{equation}
\end{widetext}

\noindent Again, the first term on the right-hand side represents the effects of domain growth.  The second term accounts for the creation of islands of zero size, initially. [$\delta(x)$ is the Dirac delta function.]  The last two terms represent the creation and annihilation of islands by coalescence, respectively.  We note that the prefactor $2v \rho_h(0, t) n(t)^{-2}$ can be obtained by writing it as $a(t)$, applying $\int_0^{\infty}dx$ to Eqs.~\ref{eq:rhoh} and~\ref{eq:rhoi}, and then comparing the two.

Unfortunately, we cannot solve Eq.~\ref{eq:rhoi} using the simple arguments that worked for $\rho_h(x,t)$.  The main difference is that a hole is created by {\it nucleation} only, while an island of nonzero size is created by growth and/or the {\it coalescence} of two or more islands.  Thus, $\rho_i(x, t)$ is given by an infinite series of probabilities for an island to contain one seed, two seeds, three seeds, and so on.  Nevertheless, we can still obtain the asymptotic behavior of $\rho_i(x,t)$ for arbitrary $I(t)$ by Laplace transforming the above evolution equation, as in~\cite{Sekimoto}.

Applying $\int_0^{\infty} dx e^{-sx}$ to Eq.~\ref{eq:rhoi}, we find
\begin{eqnarray}
\label{eq:laplace_rho}
\frac{\partial \tilde{\rho_i}(s, t)}{\partial t} &=& -2v [s + 2g(t)] \tilde{\rho_i}(s, t)\\
&&+ 2v e^{2v \int_0^t{g(t')dt'}} \cdot \tilde{\rho_i}(s, t)^2 + I(t)S(t),\nonumber
\end{eqnarray}

\noindent where $\tilde{\rho_i}(s, t) \equiv \int_0^{\infty} e^{-sx} \rho_i(x, t) dx$, with initiation conditions $\tilde{\rho_i}(s, 0) = 0$.  We can further simplify Eq.~\ref{eq:laplace_rho} by defining $\tilde{G_i}(s, t) = \exp\big[2v \int_0^t{g(t')dt'}\big] \cdot \tilde{\rho_i}(s, t)$, which then obeys
\begin{equation}
\label{eq:laplace_G}
\frac{\partial \tilde{G_i}(s, t)}{\partial t} = -2v [s + g(t)] \tilde{G_i}(s,t)+ 2v \tilde{G_i}(s,t)^2+I(t).
\end{equation}

\noindent If we write $\tilde{G_i}(s, t)$ as
\begin{equation}
\label{eq:GX}
\tilde{G_i}(s, t) = s + g(t) + \tilde{X}(s, t), 
\end{equation}
\noindent we find that $\tilde{X}(s, t)$ obeys the (nonlinear) Bernoulli equation~\cite{Boas}: 
\begin{equation}
\label{eq:Bernoulli}
\frac{\partial \tilde{X}(s, t)}{\partial t} = [s +  g(t)] \tilde{X}(s,t) + \tilde{X}(s,t)^2.
\end{equation}
\noindent Solving Eq.~\ref{eq:Bernoulli} and substituting back into Eq.~\ref{eq:GX}, we find the Laplace transform $\tilde{\rho_i}(s, t)$:
\begin{widetext}
\begin{eqnarray}
\label{eq:laplace}
\tilde{\rho_i}(s, t) &=& e^{-2v \int_0^t{g(t')dt'}} \tilde{G_i}(s,t)\nonumber\\
&=& e^{-2v \int_0^t {g(t')dt'}} \Bigg\{s+g(t) - \frac{ s \exp[2v(st + \int_0^t{g(t')dt'})] }
{ 1+2v \cdot s \int _{0}^{t}{\exp[2v(st' + \int_0^{t'}{g(t'')dt''})]}  dt'}\Bigg\}
\end{eqnarray}
\end{widetext}

We cannot perform the inverse Laplace transform of the above equation, even for the simple case of $I(t)$=const. [i.e., $g(t) \sim t$]~\cite{Sekimoto, Ben-Naim1996}.  However, from the form of denominator in Eq.~\ref{eq:laplace}, we observe that $\tilde{\rho_i}(s, t)$ has a single simple pole along the negative real-axis at $|s = s^*(t)| \ll 1$ for $t \gg 1$, regardless of the form that $g(t)$ may have.  Since the inverse Laplace transform can be written formally as the Bromwich integral in the complex-plane (i.e., as the sum of residues of the integrand~\cite{Arfken}), a standard strategy for obtaining the asymptotic expression of $\rho_i(x,t)$ for $x \gg 1$ is to expand $\tilde{\rho_i}(s, t)$ around $s^*(t)$ ($\vert s^*(t)\vert \ll 1$) to lowest order.  Following Sekimoto's approach, we define $K(s, t)$ to be the denominator in Eq.~\ref{eq:laplace}, which becomes
\[\tilde{\rho_i}(s, t) = e^{-\int_0^t {g(t')dt'}} \Big[s+g(t) - \frac{1}{2v} \frac{\partial K(s, t)}{\partial t} \frac{1}{K(s, t)}\Big],\]

\noindent Around $s=s^*(t)$, Eq.~\ref{eq:laplace} can be approximated as
\begin{eqnarray}
\label{eq:laplace_approx}
\tilde{\rho_i}(s, t) &\simeq&\frac{e^{-\int_0^t {g(t')dt'}}}{-2v} \frac{\partial K(s^*(t), t)}{\partial t} \frac{1}{\frac{\partial K(s^*(t), t)}{\partial s} [s-s^*(t)]}\nonumber \\
&=& + \frac{e^{-\int_0^t {g(t')dt'}}}{2v} \frac{ds^*(t)}{dt} \frac{1}{s-s^*(t)}
\end{eqnarray}

\noindent From Eq.~\ref{eq:laplace_approx}, we arrive at the following asymptotic expression for $\rho_i(x,t)$:
\begin{equation}
\label{eq:rhoi_asymp}
\rho_i(x, t) \simeq \frac{e^{-\int_0^t {g(t')dt'}}}{2v} \frac{ds^*(t)}{dt} e^{-|s^*(t)|\cdot x},
\end{equation}

\noindent for $x,t \gg 1$.  Now, both the prefactor and the exponent [the pole $s^*(t)$] can be obtained very easily by simple numerical methods.  On the other hand, an approximate expression for $s^*(t)$ itself can be found by first expanding $K(s,t)$ in powers of $st$ and then solving iteratively using Newton's method~\cite{Recipes}.  The result is
\begin{equation}
s^*(t) \simeq -\frac{1}{J_0} \bigg(1+\frac{J_1}{J_0^2}+\frac{4J_1^2-J_0J_2}{2J_0^4}\bigg),
\end{equation}
\noindent where
\[ J_n \equiv \int_0^t{e^{\int_0^{\tau}g(t')dt'} \tau^n d\tau}.\]
As we shall show below, Eq.~\ref{eq:rhoi_asymp} describes the behavior of $\rho_i(x,t)$ accurately for $x \gtrsim 2vt$.

\subsection{\label{sec:rhoi2i}Island-to-island distribution $\rho_{i2i}(x,t)$}

While most studies of 1D nucleation-growth have focused on $\rho_h(x, t)$ and $\rho_i(x, t)$ exclusively, the distribution of the distances between two centers of adjacent islands [the island-to-island distribution $\rho_{i2i}(x,t)$] has important applications.  For instance, whether homogeneous nucleation is a valid assumption cannot be known {\it a priori}.  Indeed, in the recent DNA replication experiment that motivated this work, the ``nucleation" sites for DNA replication along the genome were found to be not distributed randomly, a result that has important biological implications for cell-cycle regulation~\cite{Jun}.

\begin{figure}[!t]
\centering
\includegraphics[width=3.4in]{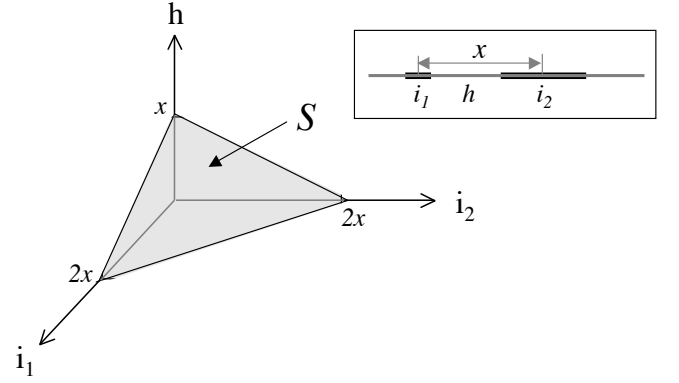}
\caption{Constraint plane $S: (i_1+i_2)/2 + h = x$.}
\label{fig:rhoi2i}
\end{figure}

In the 1D KJMA model, Sekimoto has shown that a constant nucleation function $I_0$ cannot produce correlations between domain sizes~\cite{Sekimoto}.  We speculate that the same holds true for any local nucleation function $I(x, t)$, a conclusion that is also supported by computer simulation~\cite{Jun, footnote1}.  Assuming a local nucleation function, we can write the formal expression for $\rho_{i2i}(x,t)$ directly in terms of $\rho_i(x,t)$ and $\rho_h(x,t)$:
\begin{equation}
\label{eq:rhoi2i}
\rho_{i2i}(x, t) = c \int_{\{i_1, h, i_2\} \in S} {\rho_i(i_1, t) \rho_h(h, t) \rho_i(i_2, t) dS},
\end{equation}

\noindent where $S$ designates the constraint plane shown in Fig.~\ref{fig:rhoi2i} [$S: (i_1+i_2)/2$+$h$=$x$].  The normalization coefficient $c$ can be obtained easily from the relation $\int_{0}^{\infty} \rho_{i2i}(x,t) dx = \int_{0}^{\infty} \rho_{i}(x,t) dx = \int_{0}^{\infty} \rho_{h}(x,t) dx = n(t)$.  From Eq.~\ref{eq:rhoi2i} and Fig.~\ref{fig:rhoi2i}, it is easy to see that $\int_{0}^{\infty} \rho_{i2i}(x,t)dx = c [n(t)]^3$, and therefore $c=[n(t)]^{-2}$.
Since the full solution for $\rho_i(x,t)$ is not known, we cannot integrate Eq.~\ref{eq:rhoi2i}.  However, we can still obtain an asymptotic expression for $\rho_{i2i}(x,t)$ using Eqs.~\ref{eq:sol_rhoh} and~\ref{eq:rhoi_asymp}.  For $x \gg 1$, taking into account the constraint $S$, we find
\begin{eqnarray}
\label{eq:rhoi2i_asymp}
\rho_{i2i}(x, t) &\sim& \int_{\{i_1, h, i_2\} \in S} {e^{-|s^*(t)|\cdot i_1 - g(t) \cdot h - |s^*(t)| \cdot i_2 }dS}\nonumber\\
&\sim& e^{-g(t) x} + e^{-2|s^*(t)| x} \big[-1+g(t) x - 2|s^*(t)| x\big].\nonumber\\
\end{eqnarray}

As we shall show later, the bottom Eq.~\ref{eq:rhoi2i_asymp} is an excellent approximation for all range of $x$ and time $t$.  Note that the first term on the right-hand side has the same asymptotic behavior as the hole-size distribution $\rho_h(x,t)$, while the exponential factor in the second term comes from the product of island-size distributions $\sim e^{-|s^*(t)|\cdot i_1}$ and $\sim e^{-|s^*(t)|\cdot i_2}$.  The asymptotic behavior of $\rho_{i2i}(x, t)$ is dominated by $\rho_h(x, t)$ for $f<0.5$ and by $\rho_i(x, t)$ for $f>0.5$ (see below).  But at all times, we emphasize that $\rho_{i2i}(x,t)$ is asymptotically exponential for large $x$.  From the mathematical point of view, both $\rho_i(x,t)$ and $\rho_h(x,t)$ have exponential tails at large $x$, and the integral of the product of exponential functions again produces an exponential.

\section{\label{sec:sim}Numerical simulation}

Often, one has to deal with systems for which analytical results are difficult, if not impossible, to obtain.  For example, the finite size of the system may affect its kinetics significantly, or the variation of growth velocity at different regions and/or different times could be important.  In such cases, computer simulation is the most direct and practical approach.

For one-dimensional KJMA processes, the most straightforward simulation method is to use an Ising-model-like lattice, where each lattice site is assigned either 1 or 0 (or $-1$, for the Ising model) representing island and hole, respectively.  The natural lattice size is $\Delta x = v \Delta t$, with $v$ the growth velocity.  At each timestep $\Delta t$ of the simulation, every lattice site is examined.  If 0, the site can be nucleated by the standard Monte Carlo procedure, i.e., a random number is generated and compared with the nucleation probability $I(t) \Delta x \Delta t$.  If the random number is larger than the nucleation probability, the lattice site switches from 0 to 1.  Once nucleation is done, the islands grow by $\Delta x$, namely, by one lattice size at each end.  

Although straightforward to implement, the lattice model is slow and uses more memory than necessary, as one stores information not only for the moving domain boundaries but also for the bulk.  Recently, Herrick {\it et al.} used a more efficient algorithm~\cite{Herrick}.  Specifically, they recorded the positions of moving island edges only.  Naturally, the nucleation of an island creates two new, oppositely moving boundaries, while the coalescence of an island removes the colliding boundaries.

\begin{figure}[!t]
\centering
\includegraphics[width=3.4in]{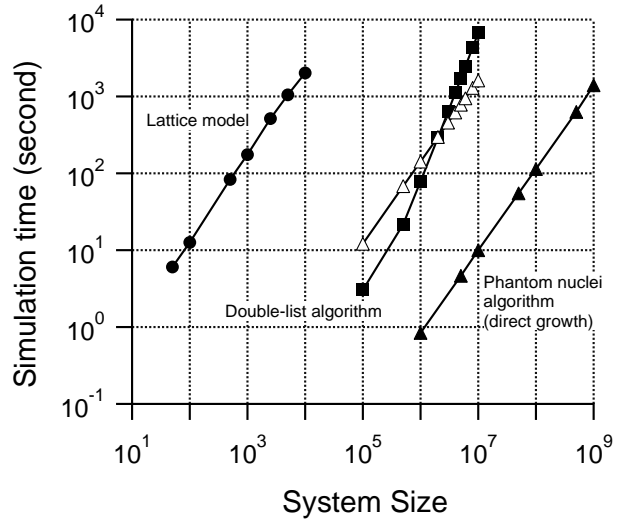}
\caption{Comparison of simulation times for the three algorithms discussed in the text.  Circles are used for the lattice-model algorithm, squares for the double-list algorithm, and triangles for the phantom-nuclei algorithm.  For each system size, the number of Monte Carlo realizations ranges from 5--20, and the lines connect the average simulation times.  The double-list algorithm is two to three orders of magnitude faster than the lattice algorithm, while the phantom-nuclei algorithm ranges from three to five orders of magnitude faster, depending on the number of time points at which one records data.  The filled triangles show the fastest case, with only one time point requested, while the open triangles show the slowest case, where data are recorded at each intermediate time step.}
\label{fig:sim_time}
\end{figure}

For the present study, we have developed two other algorithms, which have improved both simulation and analysis speeds by factors of up to $10^5$ (Fig.~\ref{fig:sim_time}).  The first of these, the ``double-list" algorithm, extends the method of Herrick {\it et al.}~\cite{Herrick}.  The second of these, the ``phantom-nuclei" algorithm, is inspired by the original work of Avrami~\cite{Avrami}.

\subsection{The Double-List Algorithm}

\begin{figure}[!t]
\centering
\includegraphics[width=3.4in]{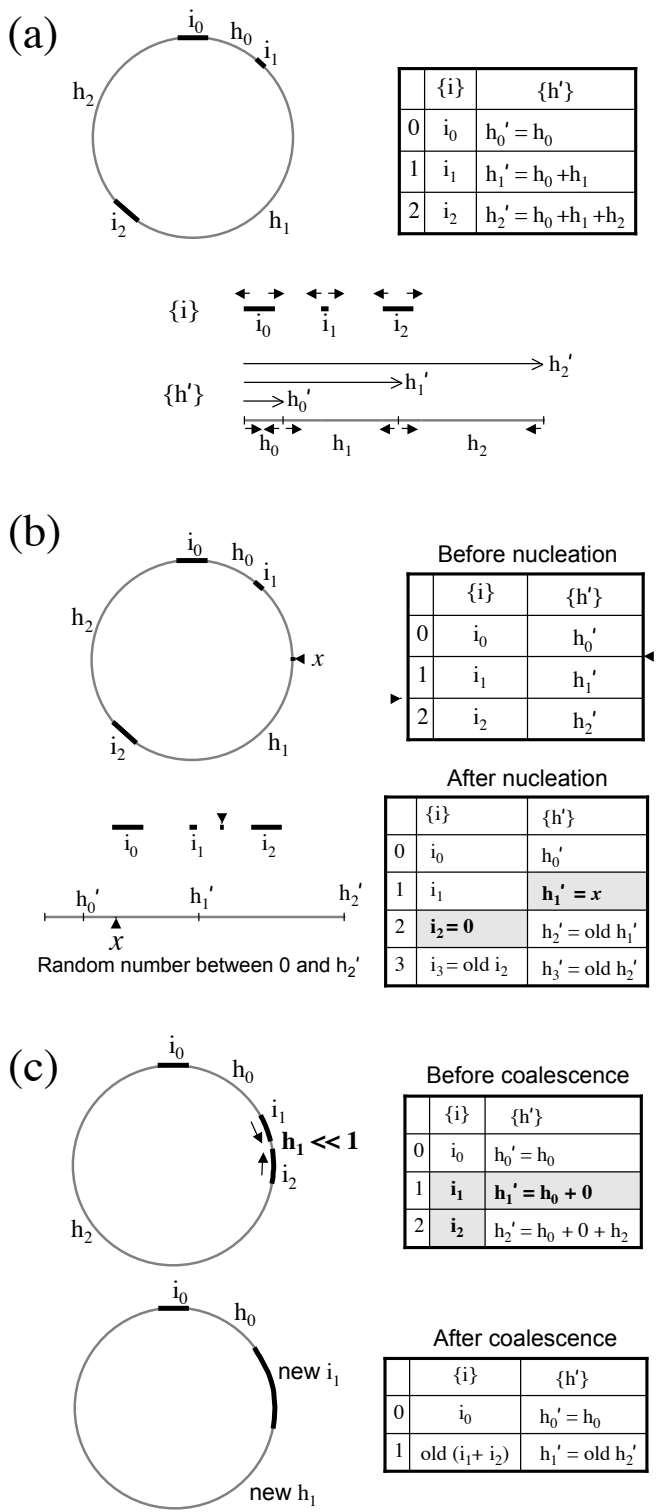}
\caption{Schematic description of the double-list algorithm.  (a) Basic set-up for lists $\{i\}$ and $\{h'\}$.  Note that $\{h'\}$ records cumulative lengths.  (b) Nucleation.  (c) Coalescence due to growth.}
\label{fig:double_list}
\end{figure}

Fig.~\ref{fig:double_list} describes schematically the double-list algorithm.  The basic idea is to maintain two separate lists of lengths: $\{i\}$ for islands, $\{h\}$ for holes~\cite{footnote_sim}.  The second list $\{h\}$ records the cumulative lengths of holes, while $\{i\}$ lists the individual island sizes.  Using cumulative hole lengths simplifies the nucleation routine dramatically.  For instance, for times $t$ ranging between $\tau$ and $\tau + \Delta \tau$, the average number of new nucleations is $\bar{N} = I(\tau) \Delta x \Delta t$.  Since the nucleation process is Poissonian, we obtain the actual number of new nucleations $N = p(\bar{N})$ from the Poisson distribution $p$.  We then generate $N$ random numbers between 0 and the total hole size, namely, the largest cumulative length of holes $h_{max}$ (the last element of $\{h\})$.  The list $\{h\}$ is then updated by inserting the $N$ generated numbers in their rank order.  Accordingly, $\{i\}$ is automatically updated by inserting zeros at the corresponding places.  If $\{h\}$ were to record the actual domain sizes as $\{i\}$ does, the nucleation routine would become much more complicated because the individual hole sizes would have to be taken into account as weighting factors in distributing the nucleation positions along the template.

Fig.~\ref{fig:sim_time} compares the running times for two different algorithms: the standard lattice model vs.~the continuous double-list algorithm described above.  We wrote and optimized both programs using the Igor Pro programming language~\cite{igorpro}, and they were run on a typical desktop computer (Pentium P3, 900 MHz).  For both, we used the same simulation conditions: timestep $\Delta t = 0.1$, nucleation rate $I(t) = 10^{-5}t$, and growth velocity $v = 0.5$.  Note that the performance of the lattice algorithm is $O(N)$, whereas the double-list algorithm is roughly $N^{1.5-2}$ for $10^5 \leq N \leq 10^7$.  The main reason is that the double-list algorithm has to maintain dynamic lists $\{i\}$ and $\{h\}$.  This requires searching and removing/inserting elements (as well as minor sorting), where each algorithm is linear, or roughly $O(N^2)$ in overall.  However, the double-list algorithm performed almost 3 orders of magnitudes faster than the lattice algorithm even at a system size of $10^7$, and we did not attempt to improve the efficiency further, for example, by using a binary search.

Finally, the relative storage requirements for the lattice algorithm compared to the double-list algorithm can be estimated by the ratio $N_{lat}/n_{max}$, where $N_{lat}$ is the total number lattice sites per unit length and $n_{max}$ is the domain density.  Equivalently, one may use $l_{min}/\Delta x$, where $l_{min}$ is the minimum island-to-island distance and $\Delta x$ the lattice size.  Since one usually sets up the simulation conditions such that $l_{min} \gg \Delta x$, the double-list algorithm requires much less memory.

\subsection{The Phantom-Nuclei Algorithm}

\begin{figure}
\centering
\includegraphics[width=2.5in]{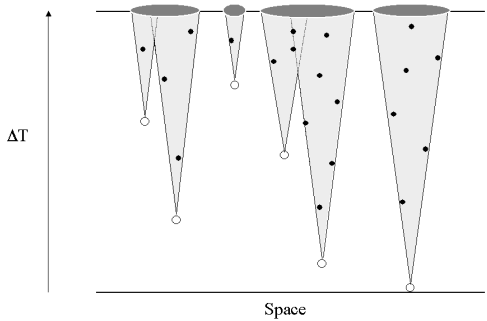}
\caption{Schematic description of the phantom-nuclei algorithm.  The figure shows the distribution of potential initiation sites in the spacetime plane.  The open circles denote sites that do initiate, while the ``phantom" filled circles, lying in the ``shadow" of the open circles, do not initiate.}
\label{fig:phantom}
\end{figure}

Fig.~\ref{fig:phantom} describes schematically the phantom-nuclei algorithm.  The basic idea is to capitalize on the ability to specify when and where in the two-dimensional spacetime plane lie potential initiation sites, in advance of the actual simulation.  Thus, in Fig.~\ref{fig:phantom}, the circles, which represent potential initiation sites, are laid down in a first part of the simulation.  One then uses an algorithm, described below, to determine which of these potential sites actually initiates (these are denoted by open circles) and which cannot fire because the system has already been transformed (these ``phantom nuclei" are denoted by closed circles).

The principal advantage of the phantom-nuclei algorithm is that one can find the state of the system at a particular time $t$ without having to calculate the system's state at intermediate time steps.  If one is interested in only a small number of system states, then the method can be significantly faster than the double-list algorithm.  The filled triangles in Fig.~\ref{fig:sim_time} illustrate a hundred-fold improvement compared to the double-list algorithm (and a $10^5$-fold improvement relative to the lattice algorithm).  On the other hand, if information is needed at every time step (or if the number of phantom nuclei is very large), the algorithm slows.  The open triangles in Fig.~\ref{fig:sim_time} show a simulation where information is collected at each time step.  The run time is comparable to the double-list algorithm.  Because there is no sorting operation, a linear time scaling is maintained.  The phantom-nuclei and double-list algorithms thus cross over at about two million sites.

The phantom-nuclei algorithm is performed as follows:
\begin{enumerate}
\item{One generates the potential initiation sites in the two-dimensional spacetime plane.  At each time, this is done in the same way as in the double-list algorithm.  The only difference is that here, the number of sites at any time is calculated over the whole length of the system (regardless of its state of transformation).  One uses two vectors to store the position and initiation time of every potential site.}

\item{One removes all initiation sites that are in positions that have already transformed before their initiation time.  (They lie in the ``shadows" in Fig.~\ref{fig:phantom}.)  Because the growth velocity is known at each time, it is straightforward to implement this.  Briefly, one first sorts the potential initiation sites by space.  Then for each potential site (indexed by $i$, with position $x_i$ and nominal initiation time $t_i$), one calculates the position of the right-hand boundary $r_i$ at the reference time point $t$.  This is given by $r_i = x_i + v (t-t_i)$ for each $i$.  One then proceeds through the list $r_i$.  If $r_i < r_j$ for any $j < i$, then site $i$ is discarded.  Finally, one repeats the analogous process moving to the left, using the left-hand boundaries $\ell_i = x_i - v (t-t_i)$.}

\item{One calculates the desired statistics at the reference time point.  This time point is arbitrary.  For the filled triangles of Fig.~\ref{fig:sim_time}, it is the last time step of the transformation process, while in the open triangles, it occurs at the next time step of the double-list simulation.  (For the latter case, the statistics were then repeatedly calculated at each time interval.)}

\end{enumerate}

In conclusion, we note that both the double-list and the phantom-nuclei algorithms are significant improvements on the more straightforward lattice algorithm.  For simple initiation schemes, where it is possible to give a function $I(t)$ for the intiation sites, the phantom-nuclei algorithm will generally be preferable.  For more complicated initiation schemes, where the initiation of sites is correlated with the activation of earlier sites, the double-list algorithm may be preferable.  In the next section, we present simulation results based on the double-list algorithm.

\section{\label{sec:compare}Comparison between theory and simulation}

\begin{figure*}[!t]
\centering
\includegraphics[width=7in]{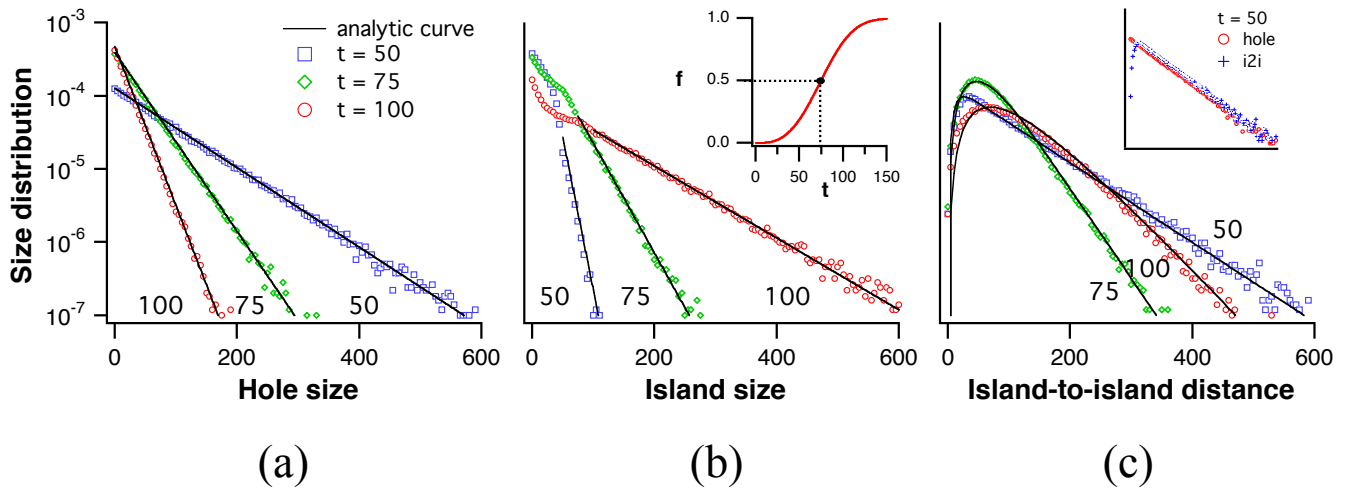}
\caption{(Color online).  Theory and simulation results for $I(t) \sim t$.  Size distributions are calculated at these timepoints: $t$ = 50, 75, and 100.  (a) Hole-size distribution $\rho_h(x,t)$.  (b) Island distribution $\rho_i(x,t)$.  The inset plots $f(t)$ vs. $t$, with the dot at $t=75$ ($f=0.5$).  (c) Island-to-island distribution $\rho_{i2i}(x,t)$.  The analytical curves have been obtained by Eq.~\ref{eq:rhoi2i_asymp}.  There is a crossover of the decay constant slightly after $t$=75 ($f=0.5$) (see text).  The inset shows $\rho_h(x,t)$ and $\rho_{i2i}(x,t)$ for $t=50$.  All figures have the same vertical range of $10^{-7} - 10^{-3}$ (log-scale).}
\label{fig:results}
\end{figure*}

In Fig.~\ref{fig:results}, we compare the various analytical results obtained in the previous sections with a Monte Carlo simulation.  Shown are $\rho_h(x,t)$, $\rho_i(x,t)$, and $\rho_{i2i}(x,t)$ for $I(t)=10^{-5} t$ at three different time points: $t$=50, 75, and 100.  The system size is $10^7$ and the growth rate is $v=0.5$.  The chosen form of accelerating $I(t)$, linear in time, is the simplest nontrivial nucleation scenario.  It is also relevant to the description of DNA replication kinetics in {\it Xenopus} early embryos, where the  $I(t)$ extracted from experimental data has a bilinear form~\cite{Herrick}.

The agreement between simulation and analytical results is excellent.  In particular, we emphasize that the analytic curves in Fig.~\ref{fig:results} are not a fit.  Note that, for $x \gg 1$, all three distributions decay exponentially as predicted by Eq.~\ref{eq:sol_rhoh},~\ref{eq:rhoi_asymp}, and~\ref{eq:rhoi2i_asymp}.  [The $\rho_h(x,t)$ distributions are simple exponentials over the entire range of $x$.]

One interesting feature of $\rho_i(x,t)$ is the inflection point in the interval $0 \leq x \leq 2vt$, where $\rho_i(x,t)$ is slightly convex.  Such behavior is even more dramatic when $I(t)$=const.~\cite{Sekimoto}, and $\rho_i(x,t)$ is strongly convex.  In other words, $\rho_i(x,t)$ increases as $x$ approaches $2vt^-$, but suddenly drops discontinuously at $x=2vt$, decaying exponentially.  This peculiar behavior of $\rho_i(x,t)$ originates from the fact that any island larger than $2vt$ must have resulted from the merger of smaller islands.  Therefore, for $x \leq 2vt$, $\rho_i(x,t)$ has an extra contribution from islands that contain only a single seed in them, which makes $\rho_i(x,t)$ deviate from a simple exponential.  Although such discontinuities are expected at every $x = n \cdot 2vt$ ($n$=1, 2, 3, $\ldots$), higher-order deviations decrease geometrically and thus are almost invisible.

Finally, the island-to-island distribution $\rho_{i2i}(x, t)$ provides important insight about the ``seed distribution" and about the spatial homogeneity of the nucleation.  Note that $\rho_{i2i}(x,t)$ is not monotonic and has a peak at $x > 0$ [see Fig.~\ref{fig:results}(c)].  This is not surprising because $\rho_{i2i}(x,t) \rightarrow 0$ as $x \rightarrow 0$ from Eq.~\ref{eq:rhoi2i}.  On the other hand, we see that $\rho_{i2i}(x,t)$ decays exponentially at large $x$, as predicted in the previous section (Eq.~\ref{eq:rhoi2i_asymp}).  In contrast to $\rho_i(x,t)$ and $\rho_h(x,t)$, however, the decay constant is not a monotonic function of time.  This can be understood as follows: at early times, the large island-to-island distances come from large holes and therefore $\rho_{i2i}(x,t) \sim \rho_h(x)$, as mentioned earlier. (The inset of Fig.~\ref{fig:results}(c) confirms this.)  However, as the island fraction $f(t)$ approaches unity, the system becomes mainly covered by large islands, and $\rho_{i2i}(x,t)$ should approach $\sim \rho_i(x, t)^2$ asymptotically (see the second term in the bottom Eq.~\ref{eq:rhoi2i_asymp}).
\begin{figure}[!t]
\centering
\includegraphics[width=2.5in]{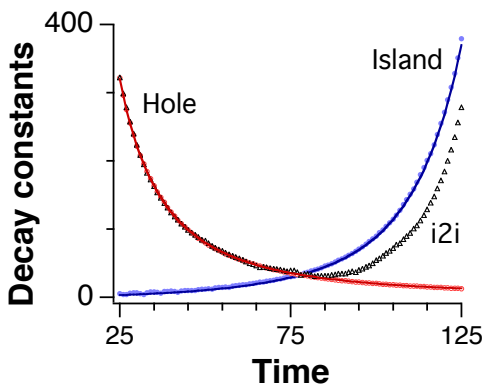}
\caption{(Color online).  Decays constants of $\rho_h(x,t)$, $\rho_i(x,t)$, and $\rho_{i2i}(x,t)$.  The symbols are simulations, and the solid lines are theory.}
\label{fig:decayC}
\end{figure}

In Fig. 8, we plot the decay constants for the three different distributions, $\tau_h$, $\tau_i$, and $\tau_{i2i}$.  Note that  when $f < 0.5$, $\tau_h \approx \tau_{i2i}$, as discussed above.  As $f \to 1$, the behavior of $\tau_{i2i}$ is controlled by $\tau_i$, as suggested by Eq. 20.  Because $\rho_{i2i} \sim \rho_i^2$, we expect $\tau_{i2i} \to 0.5 \tau_i$; however, the corrections to this relationship in Eq. 20 imply that this holds true only for large $x$ and $t$.  Note that the actual minimum of $\tau_{i2i}$ is at $f > 0.5$ because $\rho_{i2i}$ depends on $\rho_i^2$ and not $\rho_i$ alone.

One final note about the island-to-island distribution is that, unlike $\rho_i(x,t)$, it is a continuous function of $x$.  The reason for this is that for any island-to-island distance $x$, the discontinuous $\rho_i(y < x, t)$ contributes to $\rho_{i2i}(x,t)$ in a cumulative way, as can be seen in Eq.~\ref{eq:rhoi2i}.  This implies that there is no specific length scale where discontinuity can come in.  From a mathematical point of view, this is equivalent to saying that the integral of a piecewise discontinuous function (the integrand in Eq.~\ref{eq:rhoi2i}) is continuous.

\section{Conclusion}

To summarize, we have extended the KJMA model to the case where the homogeneous nucleation rate is an arbitrary function $I(t)$ of time, deriving a number of analytic results concerning the properties of various domain distributions.  We have also presented highly efficient simulation algorithms for 1D nucleation-growth problems.  Both analytical and simulation results are in excellent agreement.  In the companion paper, we discuss the application of these results to experiments in general and to the analysis of DNA replication kinetics in particular.\\

\begin{acknowledgments}
We thank Tom Chou, Massimo Fanfoni, Govind Menon, Nick Rhind, and Ken Sekimoto for helpful comments and discussions on 1D nucleation-and-growth models.  This work was supported by NSERC (Canada).
\end{acknowledgments}

\end{document}